\documentclass{article}

\usepackage{arxiv}

\usepackage[utf8]{inputenc} 
\usepackage[T1]{fontenc}    
\usepackage{hyperref}       
\usepackage{url}            
\usepackage{booktabs}       
\usepackage{amsfonts}       
\usepackage{nicefrac}       
\usepackage{microtype}      
\usepackage{lipsum}

\usepackage{xfrac}
\usepackage{amssymb}
\usepackage{amsmath}
\usepackage{hyperref}
\usepackage{soul}
\usepackage{wasysym}
\usepackage{multirow}
\usepackage{url}
\usepackage{accents}
\usepackage[super,sort&compress,comma]{natbib} 

\newcommand\blfootnote[1]{%
  \begingroup
  \renewcommand\thefootnote{}\footnote{#1}%
  \addtocounter{footnote}{-1}%
  \endgroup
}

\title{Minimising the levelised cost of electricity for bifacial solar panel arrays using Bayesian optmisation}

\author{
  Peter Tillmann,\textit{$^{a,b\dag}$}, Klaus J\"{a}ger,\textit{$^{a,b\dag}$} and Christiane Becker\textit{$^{a\ast}$}\\
}

\begin{document}
\maketitle

\begin{abstract}
Bifacial solar module technology is a quickly growing market in the photovoltaics (PV) sector. By utilising light impinging on both, front and back sides of the module, actual limitations of conventional monofacial solar modules can be overcome at almost no additional costs. Optimising large-scale bifacial solar power plants with regard to minimum levelised cost of electricity (LCOE), however, is challenging due to the vast amount of free parameters such as module inclination angle and distance, module and land costs, character of the surroundings, weather conditions and geographic position. We present a detailed illumination model for bifacial PV modules in a large PV field and calculate the annual energy yield exemplary for two locations with different climates. By applying the Bayesian optimisation algorithm we determine the global minimum of the LCOE for bifacial and monofacial PV fields at these two exemplary locations considering land costs in the model. We find that currently established design guidelines for mono- and bifacial solar farms often do not yield the minimum LCOE. Our algorithm finds solar panel configurations yielding up to  23 \% lower LCOE compared to the established configuration with the module tilt angle equal to the latitude and the module distance chosen such that no mutual shading of neighboring solar panels occurs at winter solstice. Our algorithm enables the user to extract clear design guidelines for mono- and bifacial large-scale solar power plants for most regions on Earth and further accelerates the development of competitively viable photovoltaic systems.
\end{abstract}
\blfootnote{\textit{$^{a}$~Helmholtz-Zentrum Berlin f\"ur Materialien und Energie, Albert-Einstein-Stra\ss e 16, D-12489 Berlin.}}
\blfootnote{\textit{$^{b}$~Zuse Institute Berlin, Takustra\ss e 7, D-14195 Berlin. }}
\blfootnote{$\ast$~Corresponding author; E-mail: christiane.becker@helmholtz-berlin.de.}
\blfootnote{\dag~These authors contributed equally to this work.}

\section{Introduction}     
The record power conversion efficiency (PCE) of monofacial silicon solar cells -- currently the dominant solar-cell technology -- is 26.7\%\cite{yoshikawa_silicon_2017} and approaches the physical limit of around 29.4\%, which was calculated by Richter et al.\cite{richter_reassessment_2013} Photovoltaic (PV) systems consisting of bifacial solar modules can generate a significantly higher annual energy yield (EY) than systems using conventional monofacial PV modules, because bifacial solar modules not only utilize light impinging onto their front, but also illumination onto their rear side.\cite{kopecek_towards_2018,liang_review_2019} Furthermore, advanced solar-cell concepts such as PERC, PERT, PERL (passivated emitter rear contact/totally-diffused/locally-diffused) and IBC (interdigitated back contact) can easily be manufactured as bifacial solar cells.\cite{taiyang:2018} Kopecek and Libal see bifacial solar cells as the concept with the 'highest potential to increase the output power of PV systems at the lowest additional cost'.\cite{kopecek_towards_2018} Indeed, the bifacial solar cell market has been gathering pace for a couple of years and several major PV companies, such as Sanyo,\cite{sanyo_2009_bifacial_pressrelease} Yingli,\cite{yingli_2018_bifacial_pressrelease} PVG solutions, bSolar/SolAround,\cite{bsolar_2012_bifacial_pressrelease} and Trina Solar\cite{trina_2019_bifacial_pressrelease} introduced bifacial modules. The tenth edition of the International Technology Roadmap for Photovoltaics (ITRPV) predicts a global market share of more than 50\% for bifacial modules in 2029.\cite{ITRPV2019} Large-scale bifacial PV power plants already have been realised and showed a higher energy yield than their monofacial counterparts.\cite{Ishikawa_Bifacial_2016} 

The levelized cost of electricity (LCOE) is a very relevant economic metric of a solar power plant.\cite{fertig_economic_2016} The performance of bifacial solar modules is heavily affected by their surroundings, because they can accept light from almost every direction. Hence, a vast amount of parameters influence the resulting LCOE, for example the module and land costs, module distance and inclination angle, albedo of the ground, geographical position and the weather conditions at the location of the solar farm. Liang et al. recently identified comprehensive simulation models for energy yield analysis as one of the key enabling factors.\cite{liang_review_2019} As an example, we briefly discuss how only two free parameters -- land cost and module distance -- affect the resulting LCOE, which makes it challenging to identify the sweet spot yielding a minimum LCOE: If two rows of tilted solar modules are installed close to each other, many modules can be installed per area. However, at too small distances shadowing will limit the rear side irradiance and consequently the total energy yield.\cite{appelbaum_bifacial_2016,khan_ground_2019} In contrast, putting the rows of modules far apart from each other maximizes the irradiance at the rear side and the energy yield per module. The number of modules installed per area, however, is lower and the overall energy yield of the solar farm decreases. The module inclination angle is a third free parameter, closely connected to the two aforementioned module distance and land cost, and obviously affects shadowing of neighboring solar panel rows and hence energy yield and LCOE of a bifacial solar farm, too. 

Historically, the module inclination angle was usually set to the geographical latitude of the solar farm location, and the module distance was either set to a fixed value based on experience\cite{shoukry_modelling_2016} or to the minimum module distance without mutual shadowing on the day of winter solstice at 9~am\cite{patel_worldwide_2019} or noon.\cite{kreinin_pv_2016}  However, it has turned out that these rule-of-thumb estimates often do not lead to a minimised LCOE.\cite{Grana_website_2018} One reason is that these models did not consider the cost of land. Recently Patel et al.\ considered land costs when optimising bifacial solar farms.\cite{patel_worldwide_2019} However, also in this study the module distance and inclination angle were preset according to above mentioned winter solstice rule. Considering the enormous market growth of bifacial solar cell technology, finding the optimum configuration yielding minimum LCOE is highly desired.
With the PV system costs  in \$ per Watt peak (Wp), land costs in \$ per area and the geographic location of the solar farm as known input variables, \textit{inversely} finding the optimal geometrical configuration of a bifacial PV field is a computational challenging multi-dimensional optimisation task. 

In this study, we apply a multi-parameter Bayesian optimisation in order to minimise the LCOE of large-scale bifacial solar power plants. We present a comprehensive illumination model for bifacial solar arrays and calculate the annual energy yield (EY) based on TMY3 (Typical Meteorological Year 3) data for two exemplary locations near Dallas and Seattle. We calculate optimal module inclination angles and module distances yielding minimal LCOE for various module to land cost ratios. We find that our calculated optima strongly depend on both the module to land cost ratio and the geographical location. We conclude that currently used rule-of-thumb estimates for optimal module distance and tilting angle must be reconsidered. 
Our method enables the user to extract clear design guidelines for mono- and bifacial large-scale solar power plants principally anywhere on Earth.

\section{Illumination model}
\label{sec:illumination}
\begin{figure*}
 \centering
 \includegraphics{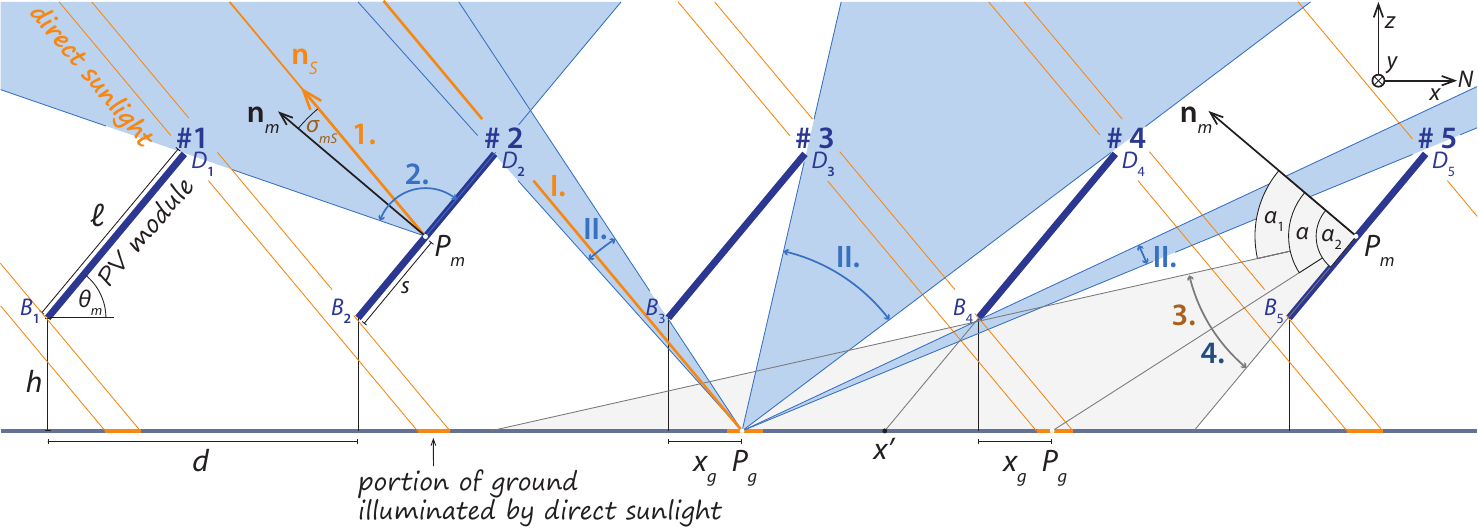}
 \caption{Illustrating the geometrical configuration of a (periodic) PV field and the illumination components, which reach each module on the front. The modules are labeled with \#1-\#5. At \#1, the geometrical parameters $h$, $\ell$, $d$ and $\theta_m$ are illustrated -- $d$ is the horizontal length of a \emph{unit cell}. At \#2, the two irradiance components illuminating the module from the sky at $P_m$ are indicated: 1.\ direct and 2.\ diffuse. Below \#3, the I.\ direct and II.\ diffuse illumination of point $P_g$ on the ground are illustrated -- here diffuse illumination origins from three angular intervals. On \#5 the angular range of light reaching $P_m$ from the ground is indicated. It consists of 3.\ direct and 4.\ diffuse light being reflected from the ground. Components 1.-4.\  are summarized in Table \ref{table:irr}. Here, we assume w.l.o.g.\ that the PV system is located on the northern hemisphere and oriented towards South.}
 \label{fig:sketch}
\end{figure*}

With the illumination model we calculate the irradiance onto a solar module, which is placed somewhere in a big PV-field. We assume this field to be so big that effects from its boundaries can be neglected. Further, we assume the modules to be homogeneous: we neglect effects from the module boundaries or module space in between the solar cells. Hence, we can treat this problem as 2-dimensional with periodic boundary conditions, as illustrated in Fig.\ \ref{fig:sketch}. A similar approach was pursued for example by Marion et al.\cite{Marion2017} In the current model we assume the solar modules to be completely black, which means they do not reflect any light which could reach another module.

The PV field is irradiated from direct sunlight under the  \emph{Direct Normal Irradiance} (DNI)\footnote{The \emph{irradiance} or intensity is the radiant power a surface receives per area.} and the direction $\mathbf{n}_S$, which is determined by the solar azimuth $\phi_S$ and the solar zenith $\theta_S$. The latter is connected to the solar altitude $a_S$ (the height above the ground) via $a_S = 90^\circ - \theta_S$. Further, the PV field receives diffuse light from the sky, which is given as \emph{Direct Horizontal Irradiance} (DHI). However, for calculating the total irradiance onto the module, also light reflected from the ground and shadowing by the other modules must be taken into account.

Figure \ref{fig:sketch} shows the different components of light, which can reach the front of a PV module at point $P_m$. The numbers 1.-4.~correspond to the numbers in the figure -- illumination on the sky is w.l.o.g.\ indicated for module \#2 while illumination from the ground is indicated w.l.o.g.\ for module \#5.
\begin{enumerate}
    \item Direct sunlight hits the modules under the direction $\mathbf{n}_S$. It leads to the irradiance component $I_{\text{dir, }f}^\text{sky}(s) = \text{DNI}\cdot\cos\sigma_{mS}$, where $s$ is the distance between the lower end of the module $B_2$ and $P_m$, $s=\overline{B_2P_m}$, and $\sigma_{mS}$ is the angle between the module surface normal and the direct incident sunlight.
    \item Diffuse skylight $I_{\text{diff, }f}^\text{sky}(s)$ hits the module at $P_m$ from directions within the wedge determined by $\sphericalangle D_1P_mD_2$. Diffuse light does not only reach the module from directions within the $xz$-plane but from a \emph{spherical wedge}, which is closely linked to the \emph{sky view factor} as for example used by Calcabrini et al.\cite{calcabrini_simplified_2019}
    \item $I_{\text{dir, }f}^\text{gr.}(s)$ denotes direct sunlight that hits the module after it was reflected from the ground. 
    \item Finally, $I_{\text{diff, }f}^\text{gr.}(s)$ denotes diffuse skylight that hits the module after it was reflected from the ground.
\end{enumerate}
All four components are summarized in Table \ref{table:irr}. Table \ref{table:input} denotes all parameters that are used as input to the model. 

\begin{table}
	\centering
	\caption{\label{table:irr} The four irradiance components which constitute the illumination of a solar module in dependence of the position $P_m$ on the module as defined in Fig.~\ref{fig:sketch}, where $s$ is the distance $\overline{BP_m}$. These components have to be considered for front and back sides -- hence eight components in total. The numbers correspond to the numbers in Fig.~\ref{fig:sketch}.}
	\begin{tabular}{lll}
	\hline
	1. & direct irradiance from the sky & $I_\text{dir}^\text{sky}(s)$ \\
	2. & diffuse irradiance from the sky  & $I_\text{diff}^\text{sky}(s)$ \\ \hline
	\multicolumn{3}{l}{diffuse irradiance from the ground\ldots}\\
	3. & \ldots originating from direct sunlight & $I_\text{dir}^\text{gr.}(s)$ \\
	4. & \ldots originating from diffuse skylight & $I_\text{diff}^\text{gr.}(s)$ \\ \hline
	\end{tabular}
\end{table}

\begin{table}[b]
	\centering
	\caption{\label{table:input} The input parameters required to calculate the different parameters of the PV system.}
	\begin{tabular}{ll}
	\hline
	\multicolumn{2}{l}{\textbf{Module parameters} (depicted in Fig.\ \ref{fig:sketch})}\\
	$\ell$ & module length (m) \\
	$w$ & module width (m) \\
	$d$ & module spacing (m) \\
	$h$ & module height above the ground (m)\\
	$\theta_m$ & module tilt angle\\ \hline
	\multicolumn{2}{l}{\textbf{Solar parameters}} \\
	DNI & direct normal irradiance (W/m$^2$)$^\P$ \\
	DHI & diffuse horizontal irradiance (W/m$^2$)$^\P$ \\
	$\theta_S$ & zenith angle of the Sun (connected to solar altitude $a_S$ via $a_S = 90^\circ-\theta_S$\\
	$\phi_S$ & azimuth of the Sun\\
	$A$ & albedo of the ground\\
	\multicolumn{2}{l}{\small \P~This parameter also can be spectral. Then, the unit would be W/(m$^2$nm).} \\ \hline
	\multicolumn{2}{l}{\textbf{Economical parameters}}\\
	$c_P$ & peak power related system costs (\$/kWp)\\ 
	$c_L$ & land consumption related costs (\$/m$^2$)\\ \hline
	\end{tabular}
\end{table}

The total irradiance (or intensity) on front is given by 
\begin{equation}
	I_f(s) = I_{\text{dir, }f}^\text{sky}(s) + I_{\text{diff, }f}^\text{sky}(s) + I_{\text{dir, }f}^\text{gr.}(s) + I_{\text{diff, }f}^\text{gr.}(s),
\end{equation}
and similar for the back side with a subscript $b$ instead of $f$. In total, we hence consider eight illumination components on our module.

As noted above, the incident light is given as DNI and DHI. The nonuniform irradiance
distribution on the module front and back surfaces has to be considered.\cite{yusufoglu_analysis_2015,kreinin_pv_2010} For the further treatment, it is therefore convenient to define unit-less \emph{geometrical distribution functions} as
\begin{align}
    \label{eq:gdf-mod}
    \iota_{\text{dir, }f}(s) &:= \frac{I_{\text{dir, }f}(s)}{\mathrm{DNI}} &\text{and}&&
    \iota_{\text{diff, }f}(s) &:= \frac{I_{\text{diff, }f}(s)}{\mathrm{DHI}}
\end{align}
for the components arising from direct sunlight and diffuse skylight, respectively. Here we omitted the superscripts ``sky'' and ``gr.''. The calculation of the components $\iota_{\text{dir, }f}^\text{gr.}(s)$ and $\iota_{\text{diff, }f}^\text{gr.}(s)$ requires the integration over geometrical distribution functions \emph{on the ground} $\gamma_\text{dir}(x_g)$ and $\gamma_\text{diff}(x_g)$, where $x_g$ is the coordinate of the point $P_g$ on the ground. 

In particular, we have
\begin{equation}
    \label{eq:gdf-mod-gr}
	\iota_f^\text{gr.}(s) = \frac{A}{2}\int_{\alpha_1(s)}^{\alpha_2}\gamma\left[x_g\left(s,\alpha\right)\right]\cos\alpha\,\text{d} \alpha,
\end{equation}
where we omitted the subscripts ``diff'' and ``dir''. The coordinate $x_g\left(s,\alpha\right)$, on which $\gamma_\text{dir}$ and $\gamma_\text{diff}$ are evaluated, is defined such that the angle between the line $\overline{P_gP_m}$ and the module normal $\mathbf{n}_m$ is equal to $\alpha$ -- the integration parameter. In Fig.~\ref{fig:sketch} the fractions of the ground, which are illuminated by direct sunlight, are marked in orange.

\begin{figure}
 \centering
 \includegraphics{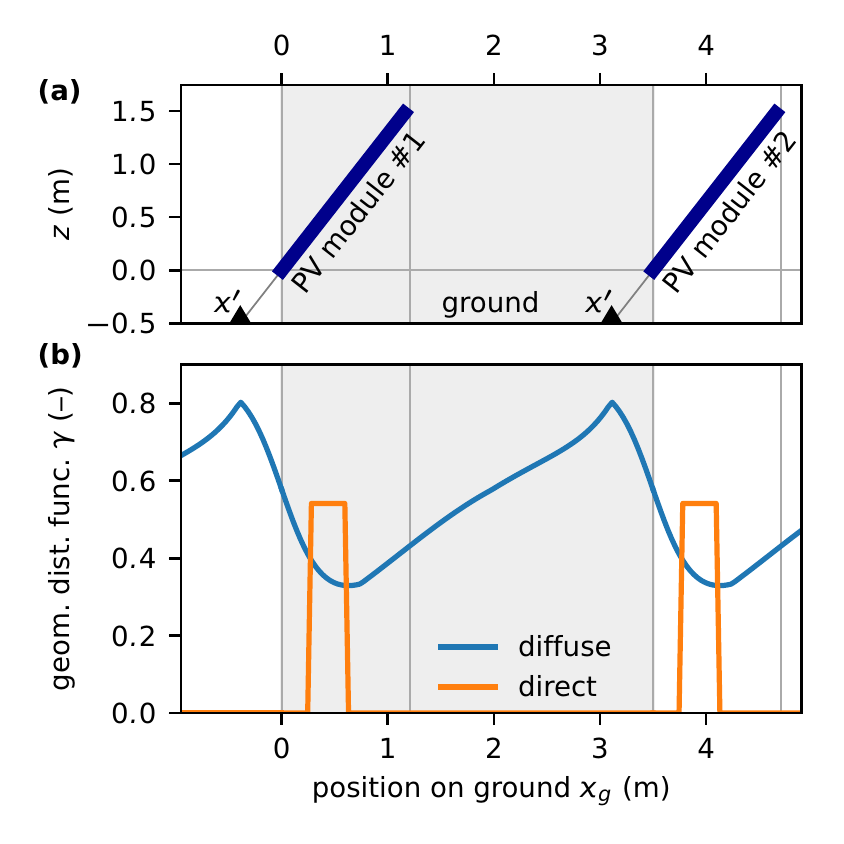}
 \caption{An example for (a) a module configuration and (b) the corresponding diffuse and direct geometrical distribution functions at the ground $\gamma_\text{diff}$ and $\gamma_\text{dir}$. The following parameters were used: $\ell = 1.96$~m, $d=3.50$~m, $h=0.50$~m and $\theta_m=52$\textdegree. The solar position for the direct component was $\theta_S=57.2$\textdegree~and $\phi_S=143.3$\textdegree~(Berlin, 20 September 2019, 11:00 CEST).  The unit cell is represented as shaded area.}
 \label{fig:G_diff}
\end{figure}

Figure \ref{fig:G_diff} shows an example for illumination onto the ground: subfigure (a) illustrates the position of the solar modules \#1 and \#2. Subfigure (b) shows the geometrical distribution functions on the ground. $\gamma_\text{diff}$ is minimal below the module where the angle covered by the module is largest; and maximal at $x'$, because here the ground sees least shadow from module \#1. 

Depending on the geometrical module parameters and the position of the Sun, the directly illuminated area (1) may lay completely within the unit cell as in the examples in Fig.~\ref{fig:sketch} and Fig.~\ref{fig:G_diff}, (2) it may extend from one unit cell into the next or (3) no direct light can reach the ground. The latter can occur when the module spacing $d$ decreases or when the solar altitude $a_S$ is low.

\begin{figure}
 \centering
 \includegraphics{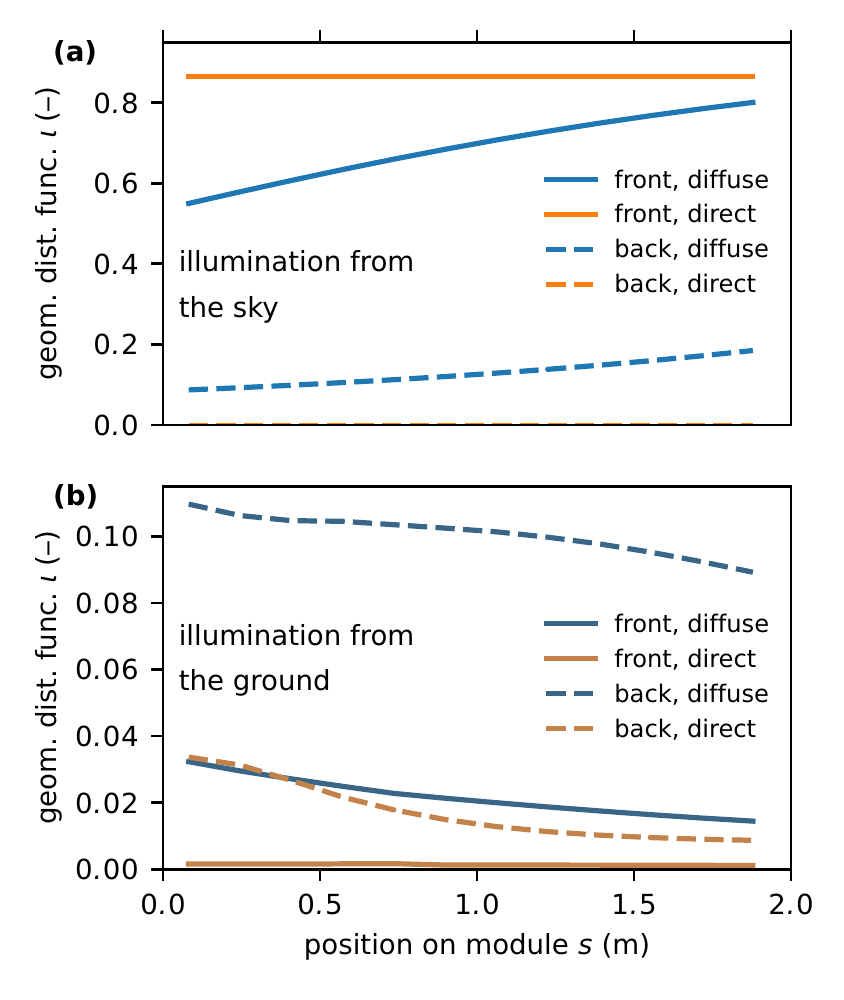}
 \caption{Geometrical distribution functions on the module for light the module receives (a) from the sky and (b) the ground. The following parameters were used: $\ell = 1.96$~m, $d=3.50$~m, $h=0.50$~m, $\theta_m=52$\textdegree, and albedo $A=30\%$. The solar position for the direct components was $\theta_S=57.2$\textdegree~and $\phi_S=143.3$\textdegree~(Berlin, 20 September 2019, 11:00 CEST).} 
 \label{fig:irr-ground}
\end{figure}

Figure \ref{fig:irr-ground} shows the eight geometrical distribution functions $\iota$ corresponding to the irradiance components hitting the PV module on its front and back sides. While the functions originating from the sky (a) are stronger on the front side, the components originating from the ground (b) are stronger on the back side. This can be understood by the opening angles: the opening angle towards the sky is larger on the front side, but the opening angle of the ground is larger at the back.

All the calculations presented in this work were performed with Python using numpy as numerical library for fast tensor operations.

\section{Annual energy yield}
\label{sec:ey-sec}
\subsection{Calculating the energy yield}
\label{sec:ey}
We calculate the \emph{annual electrical energy yield} EY by feeding the illumination model described in section~\ref{sec:illumination} with irradiance data. To demonstrate the features of the model, we use TMY3 (Typical Meteorological Year 3) data for this work. TMY3 data is well suited to estimate the solar energy yield for thousands of different locations.\cite{wilcox:2008} Amongst other parameters, the TMY3 data contain hourly DHI$(t)$ and DNI$(t)$ values. The overall EY is the sum of the energy yields harvested at the module front and back sides, $\text{EY} = \text{EY}_f + \text{EY}_b$, which are calculated with
\begin{equation}
    \label{eq:ey}
    \begin{aligned}
    \text{EY}_f=\eta_f&\left\{\sum_i   \left[\iota_{\text{dir,}f}^\text{sky}(\hat{s}_i,t_i)+\iota_{\text{dir,}f}^\text{gr.}(\hat{s}_i,t_i)\right]\text{DNI}(t_i)\Delta t\right. \\ 
    &+\left.\sum_i \left[\iota_{\text{diff,} f}^\text{sky}(\hat{s}_i)+\iota_{\text{diff,}f}^\text{gr.}(\hat{s}_i)\right]\text{DHI}(t_i)\Delta t\right\},\\
    \end{aligned}
\end{equation}
and EY$_b$ with a subscript $b$ instead of $f$. TMY3 data is available at the time stamps $t_i$ and $\Delta t$ is the time between two time stamps, which is typically 1~h for TMY3 data. $\eta_f$ and $\eta_b$ denote the power conversion efficiency for light impinging on the front and back sides of the solar module, respectively. For this work, we use $\eta_f=0.20$ and $\eta_b=0.18$, hence a \emph{bifaciality factor} of 0.9.\cite{taiyang:2018} By setting $\eta_f = \eta_b = 1$, eq.~(\ref{eq:ey}) delivers the \emph{annual radiant exposure} on the front $H_f$ (and similarly $H_b$).

The $\iota$-functions are evaluated on the position $\hat{s}_i\in\mathbb{P}_m$, where $\mathbb{P}_m=\{s_1,s_2,\ldots, s_{N_m}\}$ is the set of all considered positions along the module. In a conventional PV module, all cells are electrically connected in series and therefore the cell generating the lowest current limits the overall module current. To take this into account, we determine $\hat{s}_i$ such that 
\begin{equation}
        \left(I_f+I_b\right)(\hat{s}_i,t_i)\leqslant \left(I_f+I_b\right)(s,t_i)
\end{equation} 
for all $s\in\mathbb{P}_m$. This means that the position on the module with the lowest irradiance, which is proportional to the solar cell current, determines the overall module performance. For high-end solar modules, the module performance might be higher depending on how bypass diodes are implemented. Therefore, our condition establishes a lower bound of the module performance under certain illumination conditions. 

For the diffuse irradiation components on the module, the corresponding geometrical distribution functions $\iota^\text{diff}(s)$ need to be calculated only once because we assume the incoming diffuse radiation to be isotropic for all timestamps. For the components arising from direct sunlight, also the geometrical distribution functions $\iota^\text{dir}(s,t_i)$ are time-dependent, because they depend on the position of the Sun $(\theta_{S,i},\phi_{S,i})$,\footnote{See for example ref.~\citenum{smets:2016}, appendix E.} which we calculate using the \texttt{Python} package \texttt{Pysolar}.\cite{stafford:2018}

\subsection{Results and Discussion}

As an example, we discuss results for two locations with different climates: First, Dallas/Fort Worth area, Texas (TX), USA (Denton, 195~m elevation, 33.21\textdegree N, 97.13\textdegree W) with a \emph{humid subtropical} climate (K\"{o}ppen-Geiger classification Cfa\cite{kottek:2006}) with hot, humid summers and cool winters. Secondly, Seattle, Washington (WA), USA (Boeing Field, 47.68\textdegree N, 122.25\textdegree W) with a \emph{warm-temperate} (Mediterranean) climate (K\"{o}ppen-Geiger classification Csb\cite{kottek:2006}) with relatively dry summers and cool wet winters. Figure \ref{fig:climate} shows climate diagrams for these two locations.

\begin{figure*}
 \centering
 \includegraphics{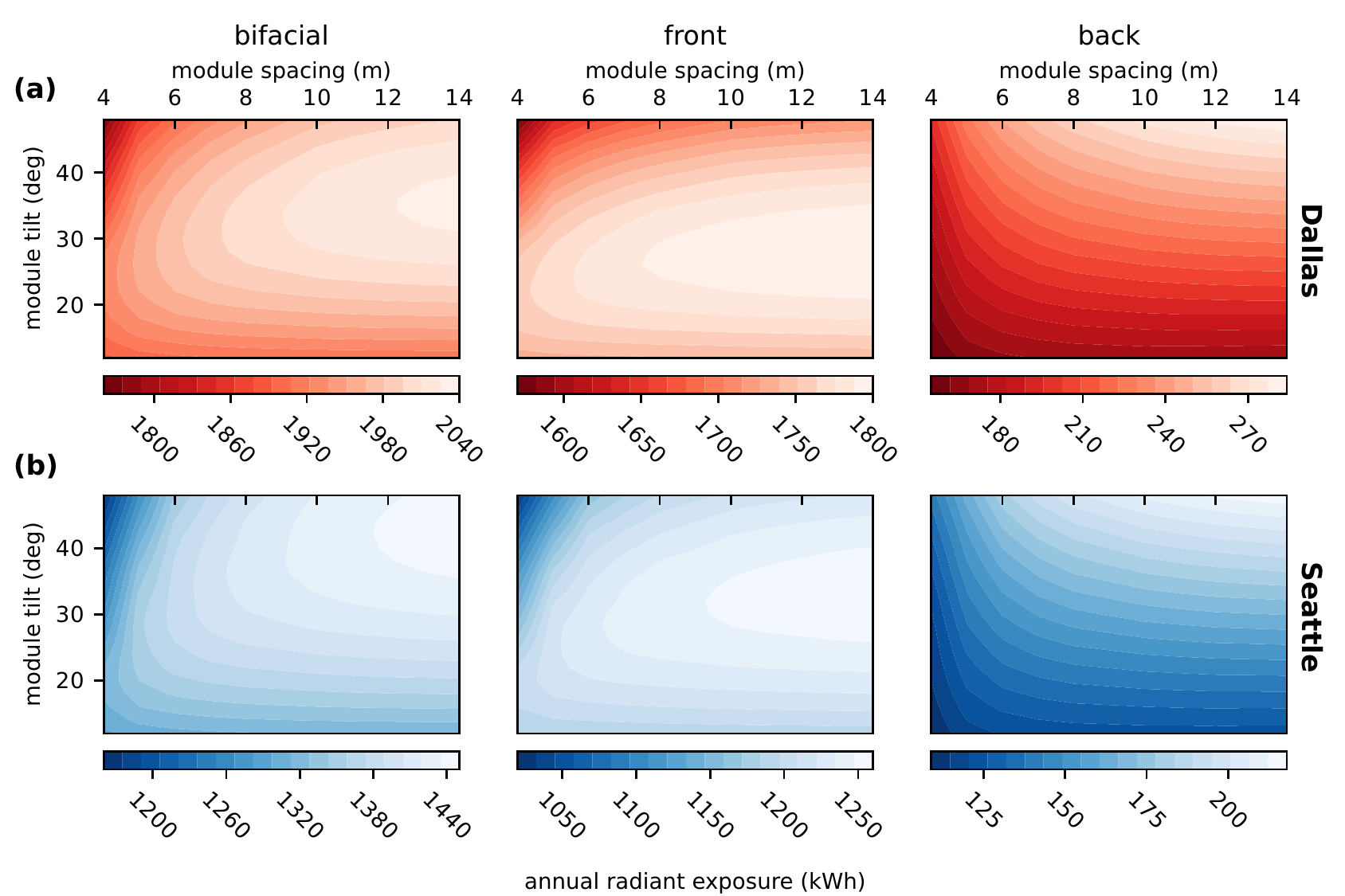}
 \caption{Annual radiant exposure for bifacial modules and the contributions from front and back sides in a large PV field as a function of module spacing $d$ and module tilt $\theta_m$. Results are shown for Dallas, TX, (top row) and Seattle, WA, (bottom row). The annual radiation yield is calculated using eq.~(\ref{eq:ey}) with $\eta_f=\eta_b=1$. Simulated with m module height $h=0.5$~m and albedo $A=30\%$.}
 \label{fig:rad-yield}
\end{figure*}

\begin{figure*}
 \centering
 \includegraphics{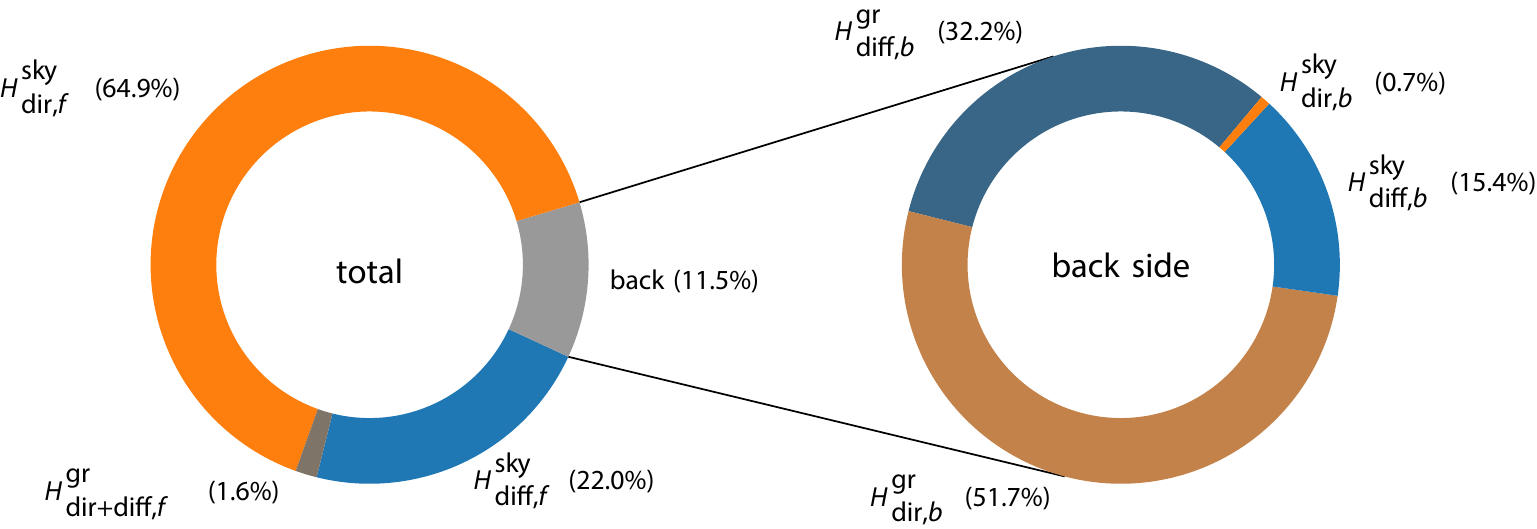}
 \caption{(left) Different annual radiant exposure components for a bifacial solar cell in Dallas.  (right) Detailed picture for the back side. Simulated with module spacing $d=10$~m, module tilt $\theta_m=34$\textdegree\ m, module height $h=0.5$~m and albedo $A=30\%$.}
 \label{fig:dallas-yield-contrib}
\end{figure*}

Figure \ref{fig:rad-yield} shows the annual radiant exposure  in (a) Dallas and (b) Seattle for bifacial PV modules (left) in a big PV field and the contributions from the front (middle) and  back sides (right). The data shown in the figure are calculated like the energy yield according to eqs.~(\ref{eq:ey}), where we set
$\eta_f=\eta_b=1$. We see that $H$ generally increases with the module spacing. However, it is not economical to have a too large distance between the rows as we will see when considering the electricity cost in Section \ref{sec:cost}.

For Dallas, the optimal angle for monofacial modules, which only can utilize front illumination, is about 28\textdegree; it is mainly determined by direct sunlight. For back illumination, $H$ increases significantly with the module inclination angle $\theta_m$: hardly any direct light reaches the module at the back, but contributions from diffuse sky and reflected from the ground increase with $\theta_m$. Increasing the module tilt further reduces the shaded area on the ground and therefore increases ground illumination. The optimal module tilt for a bifacial module is a compromise between the optimal tilt for the front and beneficial higher tilt angles for back contribution. Overall, the optimal module tilt for bifacial modules is significantly higher than for monofacial modules. Here it is about 36\textdegree.

Overall, the trends for Seattle are comparable to those for Dallas. However, we can identify differences: the overall radiant exposure is much lower because Seattle sees around 2170 annual Sun hours, compared to about 2850~h in Dallas.\cite{ozborn:2019} Further, the optimal tilt for monofacial and bifacial modules is 32\textdegree and 44\textdegree, respectively, which is explained by the higher latitude of Seattle.

For the front side illumination we see the interesting effect that, while the latitude of Seattle and Dallas differ by 14.5\textdegree, the respective optimal tilt angles only differ by 4\textdegree. This is probably because of the higher contribution on the annual radiant exposure from the summer months in Seattle compared to Dallas. While in Seattle Mai-September contribute 77 \% of the annual radiant exposure this is only 65 \% in Dallas. Because the module irradiance during the summer months (with higher elevation angles of the sun) benefits from lower tilt angle  $\theta_m$ values this can explain the difference of latitude to optimal tilt angles. The higher fraction of diffuse light in Seattle that also benefits the radiant exposure on the front side for small $\theta_m$ might additionally increase this effect.

Figure \ref{fig:dallas-yield-contrib} shows how much the different irradiation components contribute to the annual radiant exposure for a bifacial module with $d=10$~m module spacing, $\theta_m=34$\textdegree\ tilt and albedo $A = 30\%$ in Dallas: about 64\% of the total exposure arises from direct sunlight impinging onto the module front, 22\% are due to diffuse skylight impinging onto the front but the fraction of light that reaches the front from the ground is almost negligible. However, of the 12\% exposure received by the back, around 85\% is reflected from the ground. Hence, the albedo only has little influence onto the energy yield of monofacial modules but is very relevant for bifacial modules. Figure \ref{fig:seattle-yield-contrib} shows corresponding results for Seattle. Compared to Dallas, Seattle shows a nearly two per cent larger contribution by the back side. While the front side receives radiation with a ratio of nearly 3:1 of direct to diffuse light, for the back side, this ration is close to 1:1. These results show that four factors drive the gain of bifacial modules instead of monofacial modules: the albedo of the ground, the module tilt angle, the module spacing and the overall fraction of diffuse light.

 Also the mounting height $h$ affects the bifacial gain. Increasing the mounting height monotonically raises the energy yield. Therefore it is difficult to optimise this parameter without knowing additional technical and commercial constraints. However, we find that the bifacial gain starts to saturate for a height above 0.5~m, which is in agreement with work from Kreinin et al.\cite{kreinin_pv_2016} Since a mounting height of $h=0.5$~m seem realistic all simulations in our work are performed with this mounting height.

\section{Minimising the electricity cost}
\label{sec:cost}

In section \ref{sec:ey-sec} we discussed how to calculate the annual electrical energy yield and we analysed how the annual radiant exposure on the modules depends on the module spacing and tilt for two examples: Dallas and Seattle. In this section, we are going to derive a simple model for the electricity cost and perform some cost optimisations.

\subsection{Levelied cost of electricity}

As a measure for the electricity cost we use the \emph{levelized cost of electricity} (LCOE), which is a key metric for electricity generation facilities. In the simplest case, the LCOE is given as the total cost $C_F$ spent in the facility during its lifetime $T$ (in years) divided by the total amount of electric energy $E_\text{total}$ generated in that time,
\begin{equation}
    \label{eq:lcoe}
    \text{LCOE} = \frac{C_F}{E_\text{total}}=\frac{C_F}{E_F \cdot T},
\end{equation}
where $E_F$ is the electric energy generated by the PV field in one year. In more involved models also costs of capital and discount rates are taken into account.\footnote{See for example Ref.~\citenum{smets:2016}, Chapter 21.}

The total cost can be split into two components, associated with the peak power $C_P$ (including modules, inverters, mounting etc.) and the land consumption $C_L$ (lease, fences, cables etc.) of the facility.
\begin{equation}
    C_F = C_P + C_L.
\end{equation}
By considering a facility with a PV-field of $M$ rows with $N$ modules each the costs can be calculated per unit cell,
\begin{equation}
    C_F = (C_{P, m} + C_{L,m})\cdot M N.
\end{equation}

The peak-power related costs per module $C_{P, m}$ are calculated with
\begin{equation}
    C_{P, m} = c_P\cdot \eta_f\cdot I_P \cdot \ell w
\end{equation}
where $c_P$ denotes the peak-power associated costs given in $[c_{p}]=\$/\text{kWp}$, which we use as input parameter. $I_P = 1$~kW/m$^2$ is the peak irradiance as used for standardized PV characterization \cite{1992ISO9845-1:1992} and $w$ and $\ell$ denote the module width and length, respectively. $\eta_f$ denotes the power conversion efficency on the front side of the solar cell.

The cost of land consumption per module depends on module width $w$ and spacing $d$,
\begin{equation}
    \label{eq:lcm}
    C_{L, m} = c_L \cdot dw 
\end{equation}
with the land cost $c_L$ given in $[c_L] = \$/\text{m}^2$, which is an input parameter.

The annual generated electric energy of the PV field is given by
\begin{equation}
    \label{eq:EF}
    E_F = \text{EY} \cdot \ell w \cdot MN, 
\end{equation}
with the annual yield EY according to eq.~\ref{eq:ey}.

 Combining eqs.~(\ref{eq:lcoe})-(\ref{eq:EF}) and simplifying leads to the expression
\begin{equation}
    \text{LCOE} = \frac{\ell I_P\eta_f c_P + dc_L}{\ell\cdot\text{EY}\cdot T},
\end{equation}
which is independent of the field dimensions $M$  and $N$ and the module width $w$.

In this study, we assume for the overall costs of the PV system $c_m = 1000$~\$/kWp, which includes all costs over the lifetime of the solar park, such as PV module investment, balance of system cost, planning, capital cost and others. The land cost is not included in this quantity. The lifetime is assumed to be $T=25$~years, a typical time span for the power warranty of solar cell modules.\cite{smets:2016}

\begin{table}
	\centering
	\caption{\label{table:cost} Overview of used cost scenarios. Right column shows the share of land consumption on total costs for different scenarios assuming row spacing d = 10 m, module length $\ell$ = 1.96 m and $\eta_f$ = 20\%.}
	\begin{tabular}{ccc}
	$c_P$ (\$/kWp) & $c_L$ (\$/m$^2$) & $C_L/C_F$ (\%)\\
	\hline
	1000 & \phantom{0}1.00 & \phantom{0}2.5 \\
	1000 & \phantom{0}2.50 & \phantom{0}6.0 \\
	1000 & \phantom{0}5.00 & 11.3 \\
	1000 & 10.00 & 20.3 \\
	1000 & 20.00 & 33.8 \\
	\hline
	\end{tabular}
\end{table}

In our optimization, we aim to minimize the LCOE as  parameter of the module spacing $d$ and the solar module tilt $\theta_m$. We perform the optimization for five land-cost scenarios $c_L$, in which we assume to include all costs that are related to an increase of area such as lease, cables, fences etc. Table \ref{table:cost} gives an overview of the cost scenarios and the resulting fraction of the land costs  on the total costs, ($C_L/C_F$).

\subsection{Optimisation method}

As optimisation method we use Bayesian optimisation, which is well suited to find a global minimum of black box functions, which are expensive to evaluate.\cite{Shahriari2016TakingOptimization} Bayesian optimisation has been used in a wide variety of applications such as robotics,\cite{Cully2015RobotsAnimals} hyper parameter tuning\cite{Snoek2012PracticalAlgorithms} or physical systems.\cite{schneider:2017,Herbol2018EfficientOptimization}

In principle, Bayesian optimisation consists of two components: a surrogate model that approximates the black box function and its uncertainty (based on previously evaluated data points) and an acquisition function that determines the next query point from the surrogate model. After evaluating the function for the queried data point the surrogate model is updated and the next step can be computed with the  acquisition function. This cycle is repeated until a specified number of steps is reached or a convergences criteria is reached. We use the implementation from \textit{scikit-optimize} with Gaussian process as surrogate model and expected improvement as acquisition function.\cite{head:2018}

\subsection{Optimization results}
Traditionally, the optimal tilt and module spacing are often estimated with the \emph{winter solstice rule}.\cite{Patel2019AFarms, Sanchez-Carbajal2019OptimumFactors} The optimal distance between two rows of modules is defined as the shortest distance for which the shadow of a row of modules does not hit the next row of modules in a specified solar time window (e.g. 9~am - 3~pm) on winter solstice. As a rule of thumb the tilt is often chosen to be equivalent to the latitude of the facility location. However these rules do not consider the economic trade off between land costs and energy yield or typical weather patterns (e.g. foggy winters) that vary for different locations.

\begin{figure*}
 \centering
 \includegraphics{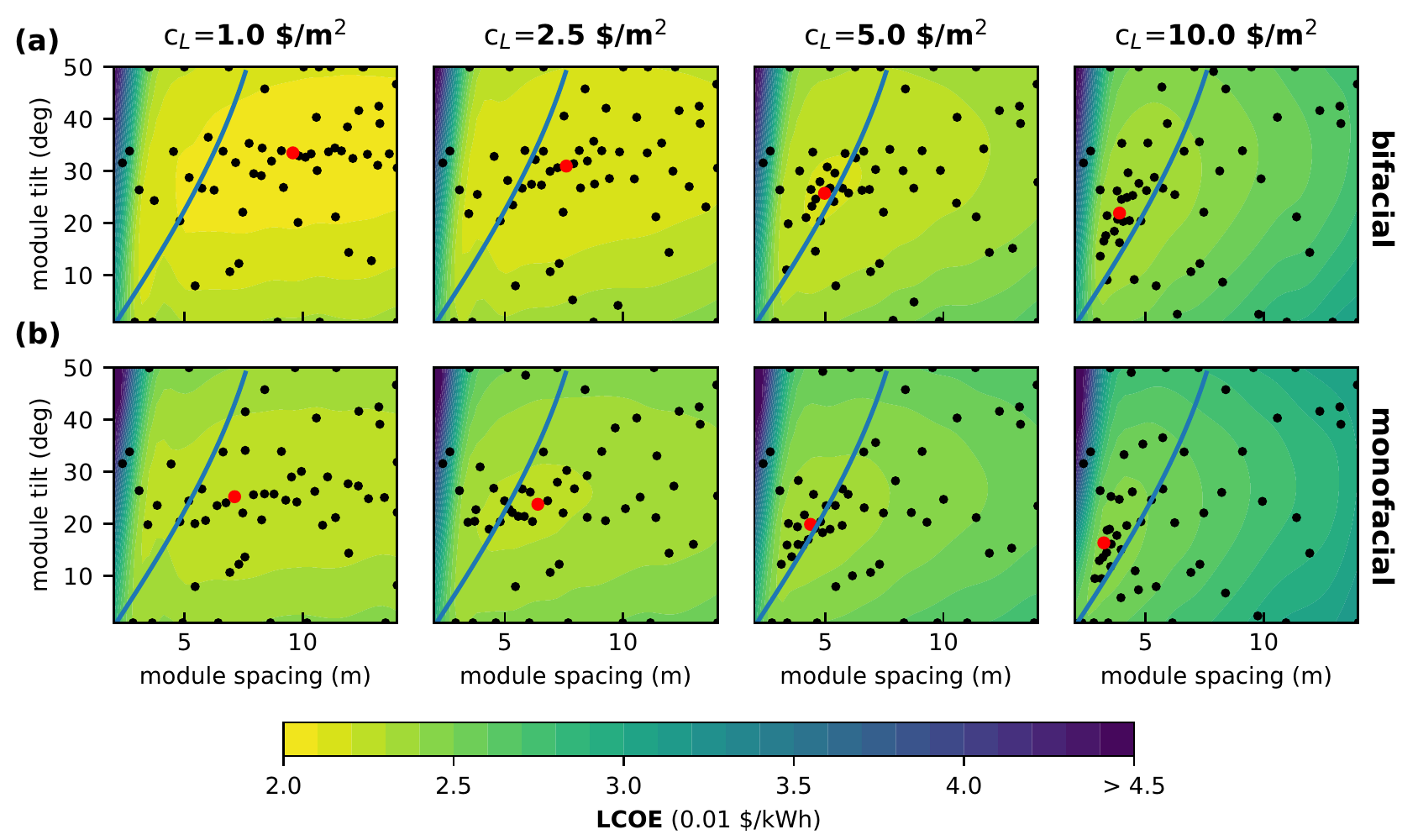}
 \caption{Results of the Bayesian optimisation for minimising LCOE of (a) \emph{bifacial} and (b) \emph{monofacial} PV modules in Dallas with the land cost $c_L$ scenarios 1, 2.5, 5, and 10 \$/m$^2$. Black dots mark evaluated configurations and the color map corresponds to the interpolation by a Gaussian process. The red dot indicates the minimal LCOE found by the optimization. The blue curves indicate rule-of-thumb module distance according to 'no shadowing of neighboring modules at winter solstice'. Simulations with albedo $A=30$~\%, module height $h=0.5$~m and peak power costs $c_p=1000$~\$/kWh.}
 \label{fig:opt-dallas}
\end{figure*}

\begin{figure*}
 \centering
 \includegraphics{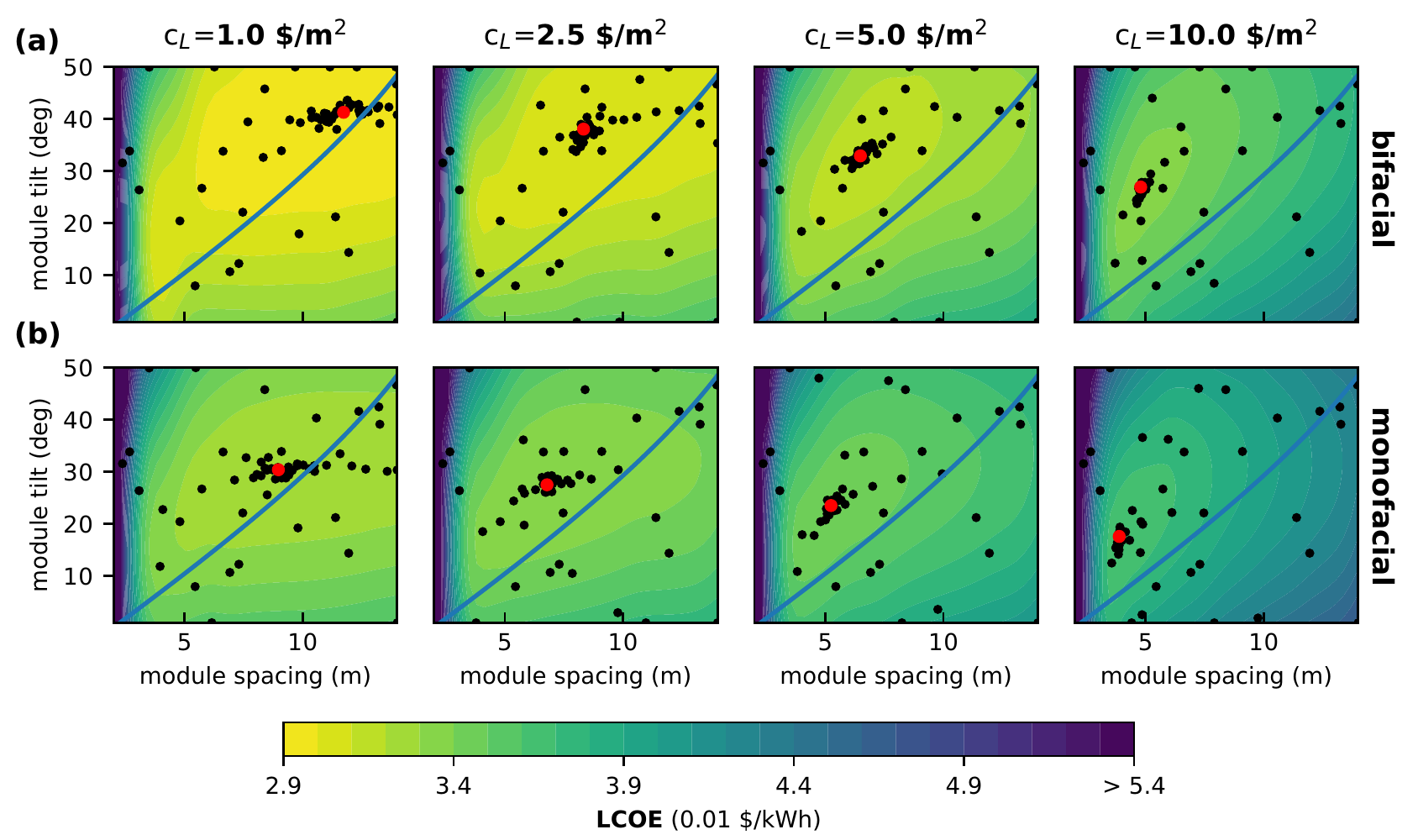}
 \caption{Results of the Bayesian optimisation for minimising LCOE of (a) \emph{bifacial} and (b) \emph{monofacial} PV modules in Seattle with the land cost $c_L$ scenarios 1, 2.5 , 5 and 10 \$/m$^2$. Black dots mark evaluated configurations and the color map corresponds to the interpolation by a Gaussian process. The red dot indicates the minimal LCOE found by the optimisation. The blue curves indicate rule-of-thumb module distance according to 'no shadowing of neighboring modules at winter solstice'. Simulations with albedo $A=30$~\%, module height $h=0.5$~m and peak power costs $c_p=1000$~\$/kWh.}
 \label{fig:opt-seattle}
\end{figure*}

Figures \ref{fig:opt-dallas} and \ref{fig:opt-seattle} shows the optimisation results for a field of (a) bifacial and (b) monofacial PV modules in Dallas and Seattle, respectively. Black dots mark evaluated data points, the red dot marks the found optimum and the color map shows the interpolation of the LCOE by the Gaussian process. The blue line indicates the winter solstice rule (9 am).

We see that the optimum shifts to smaller module spacing with increasing land cost. Further, also the optimal module tilt decreases in order to compensate for increased shadowing because of less module spacing. Overall, bifacial installations show large module spacing and higher tilt angles in optimal configurations compared to monofacial technology. With increasing land costs and therefore reduced optimal module spacing the cost landscape gets increasingly steep. The sensitivity of the optimized parameters increases and using non-optimal geometrical configurations results in increasing yield loss. Seattle shows the same trends for optimal configuration in different cost scenarios. Compared to Dallas optimal tilt and spacing are higher.

\begin{table}
	\centering
	\caption{\label{table:opt-res2} Comparing LCOE results for bifacial modules with optimized tilt and distance vs.\ rule-of-thumb parameters (module tilt equal latitude and distance according to 9 am winter solstice rule) for Seattle and Dallas. Simulations with albedo $A = 30 \%$, module height $h = 0.5$~m and peak power costs $c_P=1000$~\$/kWh.}
    \begin{tabular}{c|cr|rr|rr}
        & \multicolumn{4}{c|}{LCOE (cents)} & & \\
        $c_L$ (\$/m$^2$) &  \multicolumn{2}{c|}{optimised} &  \multicolumn{2}{c|}{rule-of-thumb} & \multicolumn{2}{c}{reduction (\%)}\\
         & \textsc{dall.} & \textsc{seat.} &  \textsc{dall.} & \textsc{seat.}& \textsc{dall.} & \textsc{seat.}\\ \hline
        \phantom{0}1.00 & 2.05 & 2.90 & 2.07 & 2.91 & \textbf{1.0} & \textbf{0.3}\\
        \phantom{0}2.50 & 2.11 & 3.01 & 2.12 & 3.06 & \textbf{0.5} & \textbf{1.6}\\
        \phantom{0}5.00 & 2.19 & 3.15 & 2.20 & 3.31 & \textbf{0.5} & \textbf{4.8}\\
                  10.00 & 2.30 & 3.35 & 2.36 & 3.81 & \textbf{2.5} & \textbf{12.1}\\
                  20.00 & 2.50 & 3.68 & 2.69 & 4.81 & \textbf{7.4} & \textbf{23.5}\\
    \end{tabular}
\end{table}

Our optimisation results differ significantly from the geometric parameters obtained from the winter solstice rule. For Dallas the winter solstice rule only provides comparable optimal parameters for $c_L = 5 \$/\text{m}^2$. In Seattle, the optimal distances are shorter and the optimal module tilts are larger than expected from the winter solstice rule for all cost scenarios. This can be understood when considering the large share of diffuse light during the Seattle winter, which mitigates shading losses significantly.

Table \ref{table:opt-res2} compares the LCOE obtained our from optimisation to results for rule-of-thumb geometries (tilt angle = latitude, distance according to 9 am winter solstice rule) for differnt land cost scenarios. Depending on the location and cost scenario we see a reduction of LCOE of up to 23\%. The rule-of-thumb approach shows its weakness especially in Seattle at high cost scenarios, where the cost landscape is very steep (see Figures \ref{fig:opt-dallas} and \ref{fig:opt-seattle}).

From these results it is clear that the winter solstice rule is not able to properly reflect different economic trade-offs or different illumination conditions over the course of the year. This is especially true when setting the tilt angle to the latitude of the location.  For a minimal LCOE module tilt and spacing should be optimised independently from each other. Further, typical weather patterns and the economic situation location must be taken into account.

\subsection{Discussion}

\begin{table}
	\centering
	\caption{\label{table:opt-res1}
	{Fraction of land cost ($C_L/C_F$), module distance $d$ and bifacial gain for optimised configurations in different cost scenarios. Simulations with albedo $A = 30 \%$, module height $h = 0.5$~m and peak power costs $c_P = 1000$~\$/kWh.}}
    \begin{tabular}{c|cr|rr|rr}
        \multirow{2}*{$c_L$ (\$/m$^2$)} &  \multicolumn{2}{c|}{$C_L/C_F$ (\%)} &  \multicolumn{2}{c|}{$d$ (m)} & \multicolumn{2}{c}{bif.\ gain (\%)}\\
         & \textsc{dall.} & \textsc{seat.} &  \textsc{dall.} & \textsc{seat.}& \textsc{dall.} & \textsc{seat.}\\ \hline
        \phantom{0}1.00 & 2.4 & 2.9 & 9.6 & 11.7 & 12.9& 15.6 \\
        \phantom{0}2.50 & 4.6 & 5.0 & 7.6 & 8.3 & 12.1& 14.4 \\
        \phantom{0}5.00 & 6.0 & 7.6 & 7.2 & 6.5 & 10.5& 12.8 \\
        10.00 & 9.0 & 10.9 & 3.9 & 4.8 & 9.3 & 10.9 \\
        20.00 & 13.9 & 15.6 & 3.2 & 3.6 & 7.6 & 8.8 \\
    \end{tabular}
\end{table}

\begin{figure*}
 \centering
 \includegraphics{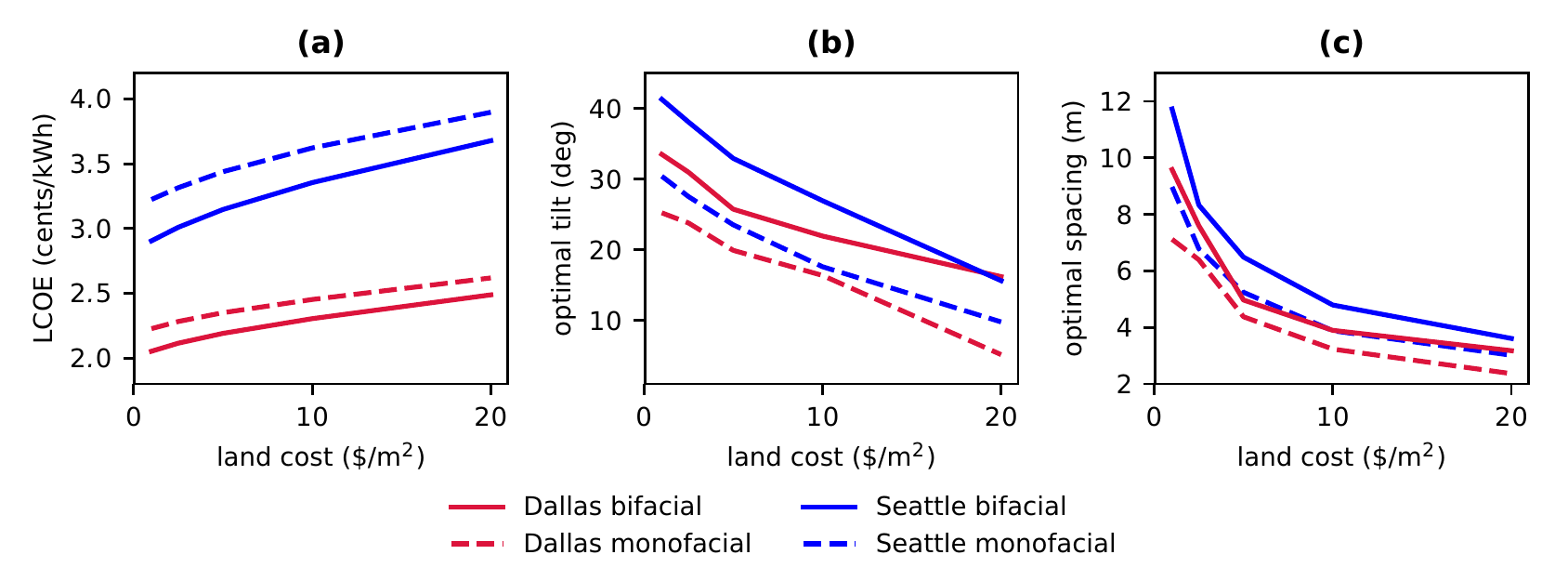}
 \caption{Results of the optimisation for different land cost scenarios in Dallas (red lines) and Seattle (blue lines): (a) lowest LCOE and (b) optimal module tilt and (c) optimal spacing. Simulations with albedo A = 30 \%, module height h = 0.5 m and peak power costs c$_p$ 1000 \$/kWh.}
 \label{fig:opt-lcoe-tilt}
\end{figure*}

The results of all optimisations are summarised in Fig.~\ref{fig:opt-lcoe-tilt} and Table \ref{table:opt-res1}. We see that the optimal LCOE increases slightly with the land cost. Further, in Seattle the LCOE difference between mono- and bifacial modules is larger as in Dallas. This is caused by the larger module tilt and diffuse light share in Seattle, which increases the fraction of illumination at the module back. As discussed above, the optimal module tilt decreases with increased land consumption cost $c_L$.

In general, we see that for a utility scale solar cell plant both, the module tilt and the distance between rows, affect the annual energy yield. Increasing the distance increases the energy yield and the costs per module  while tilt can be optimized cost-neutral. The optimal distance between rows is a compromise between increasing costs with higher land use for higher distances and lower energy yield due to shading for lower distances. This is true also for monofacial modules but due to the increased relevance of light reflected from the ground it is more relevant for bifacial modules.

The optimal configuration for bifacial solar cells depends on the radiation conditions and the albedo of the facility location. With increasing latitude (and therefore lower solar elevation angles), albedo and diffuse light contribution the bifacial gain will be increased and therefore make this type of PV technology more attractive for utility scale developers.

Cost optimisations for PV installation are quickly outdated because PV module prices have been decreasing for many years and land cost is very volatile. However the optimal installation geometry only depends on the ratio of land cost related to total costs and not absolute values. Hence, at a scenario of $c_L = 10$~\$/m$^2$ and $c_P = 1500$~\$/kWp yields the same optimization result as $c_L = 5$~\$/m$^2$ and $c_P = 750$~\$/kWp.

\section{Conclusions and Outlook}

We developed a detailed model to calculate the irradiation onto both sides of a PV module, which is located in a large PV field. With this model, we could estimate the annual energy yield for monofacial and bifacial PV modules as a function of the module spacing and the module tilt. A simple model to calculate the levelized cost of electricity combined with a Bayesian optimisation algorithm allowed us to minimise the LCOE as a function of module spacing and module tilt for different land consumption costs.

Our results show that the bifacial gain and optimal geometry depend on the specific location and cost scenario. The bifacial gain can be expected to increase for locations with higher latitude and higher diffuse light share.

The usually used rule of thumb, no shadowing at winter-solstice and module tilt angle equal to the geographical latitude, leads to suboptimal module spacing and tilt combinations, because it does not account for economic trade-offs and the influence of the local climate.
In contrast, optimising the parameters in Seattle can lead to a 23\% reduction of LCOE for high land cost scenarios. This shows the significance of site-specific optimisation and helps users to identify the configurations yielding minimal LCOE.



\section*{Conflicts of interest}
There are no conflicts to declare.

\section*{Acknowledgements}
We thank Lev Kreinin and Asher Karsenti from SolAround for fruitful discussions regarding the illumination model for bifacial solar cells.  P.\ T.\ thanks the Helmholtz Einstein International Berlin Research School in Data Science (HEIBRiDS) for funding. The results were obtained at the Berlin Joint Lab for Optical Simulations for Energy Research (BerOSE) of Helmholtz-Zentrum Berlin f\"{u}r Materialien und Energie, Zuse Institute Berlin and Freie Universit\"{a}t Berlin.



\bibliography{library.bib,mendeley.bib}
\bibliographystyle{unsrt}

\appendix
\newpage
\section*{Supporting Information}

\section{Climate diagrams}
\begin{figure*}[h]
    \centering
    \includegraphics{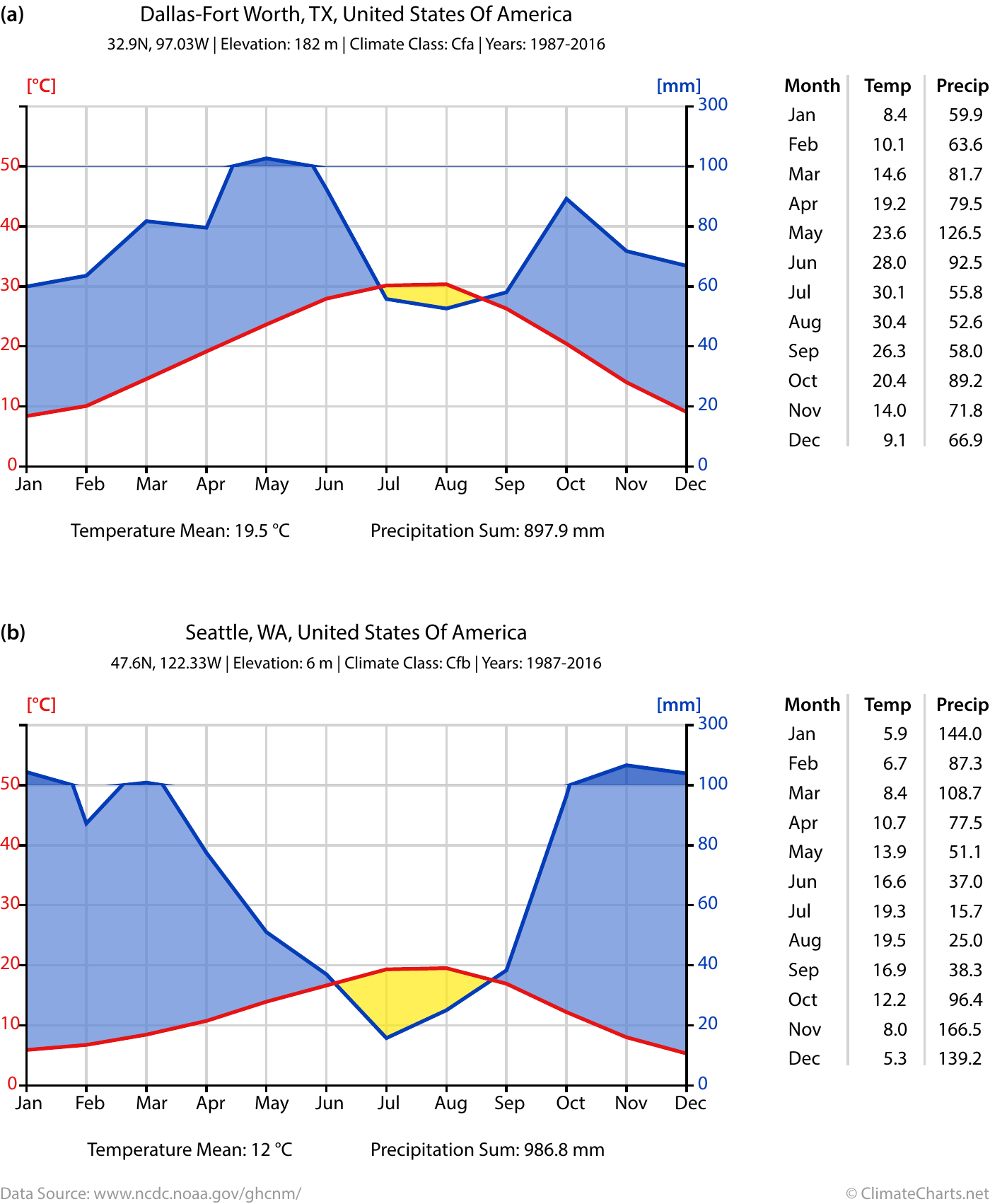}
    
    \caption{Climate diagrams for (a) the Dallas/Fort Worth area and (b) Seattle. These charts were generated on \url{ClimateCharts.net} and are  licensed under a Creative Commons Attribution 4.0 International License.}
    \label{fig:climate}
\end{figure*}

Figure \ref{fig:climate} shows climate diagrams for (a) the Dallas/Fort Worth area and (b) Seattle. Depending on the source, the climate for Seattle is classified as warm temperate (Csb) or as oceanic (Cfb). 

\section{Irradiance distribution}
\begin{figure*}
    \centering
    \includegraphics{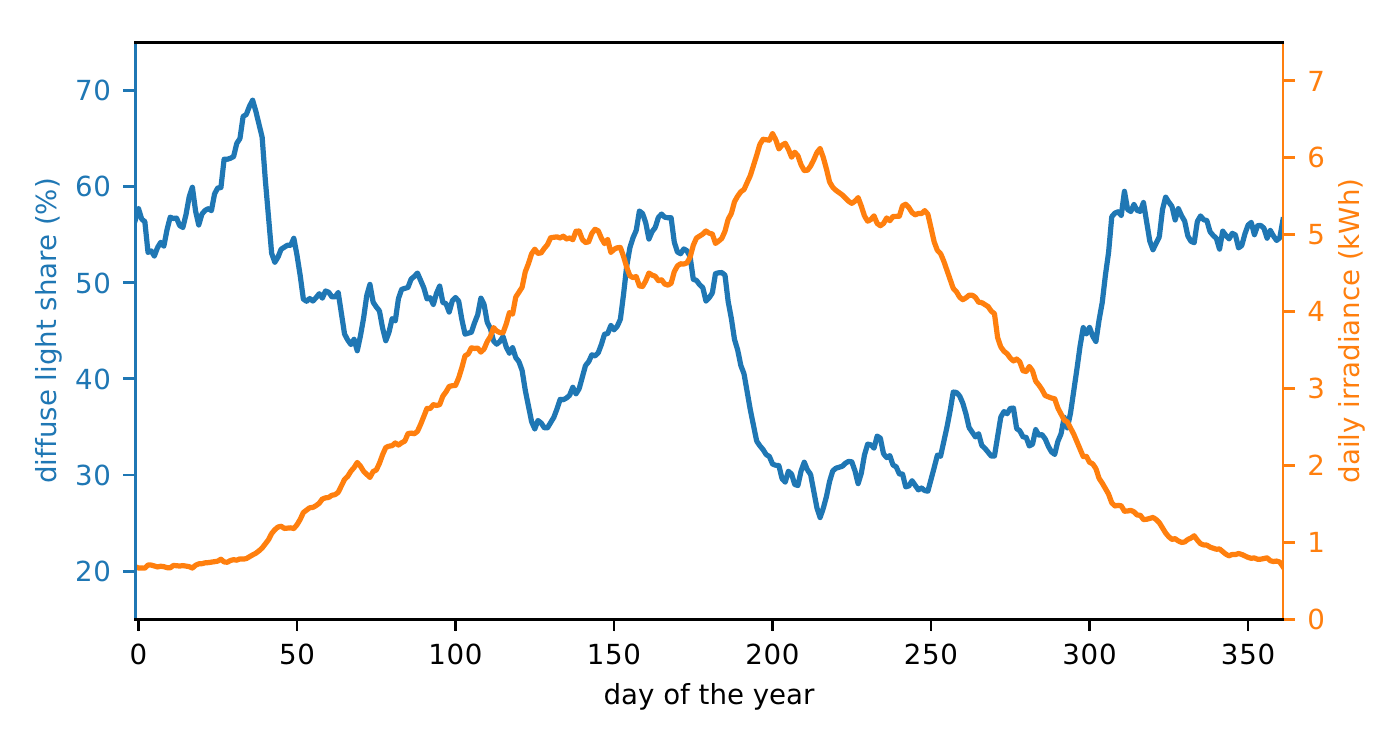}
    \caption{30 days moving average of diffuse light share (blue curve) and daily global horizontal irradiance (orange curve) for the typical meteorological year in Seattle}
    \label{fig:dist-seattle}
\end{figure*}

\begin{figure*}
    \centering
    \includegraphics{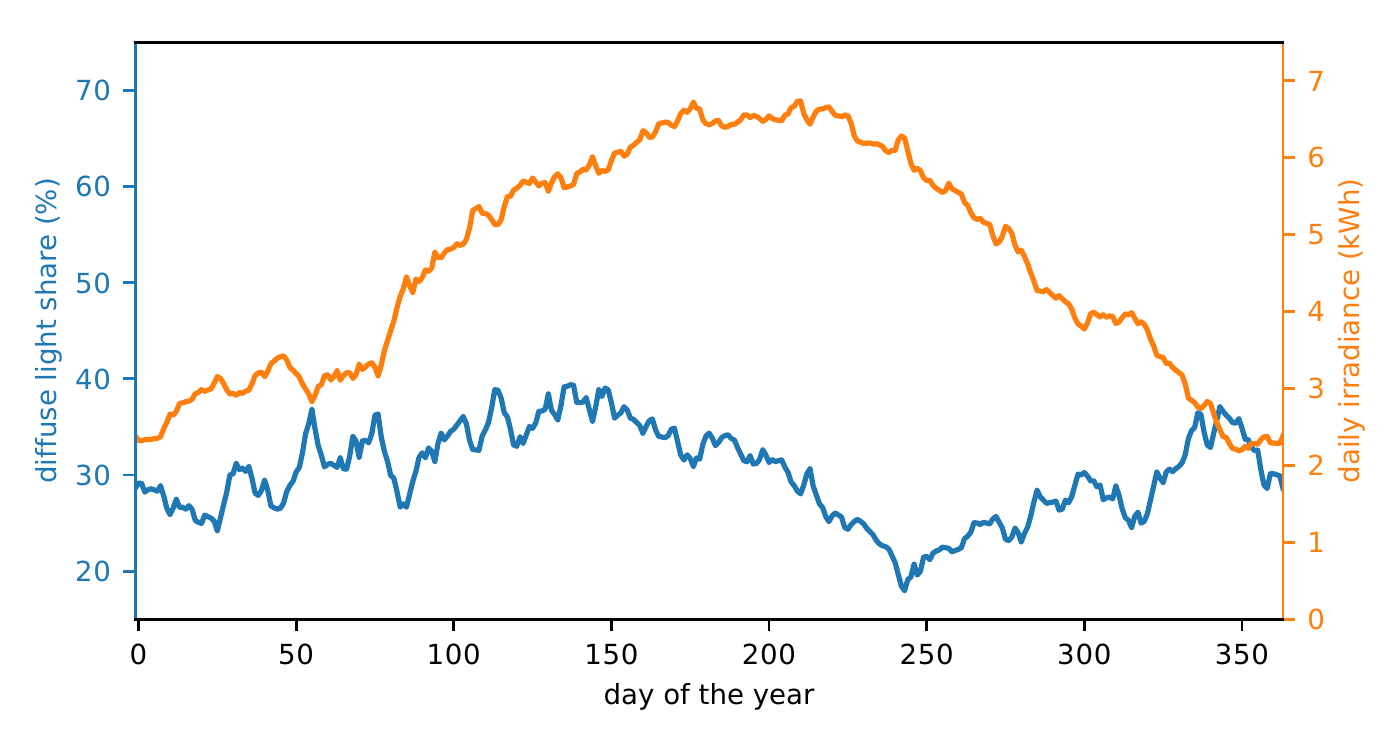}
    \caption{30 days moving average of diffuse light share (blue curve) and daily global horizontal irradiance (orange curve) for the typical meteorological year in Dallas}
    \label{fig:dist-dallas}
\end{figure*}

Figure \ref{fig:dist-seattle} and Figure \ref{fig:dist-dallas} show the distribution of the global horizontal irradiance and the share of diffuse light over the course of the year for Seattle and Dallas.

\section{Further results}
\begin{figure*}
 \centering
 \includegraphics{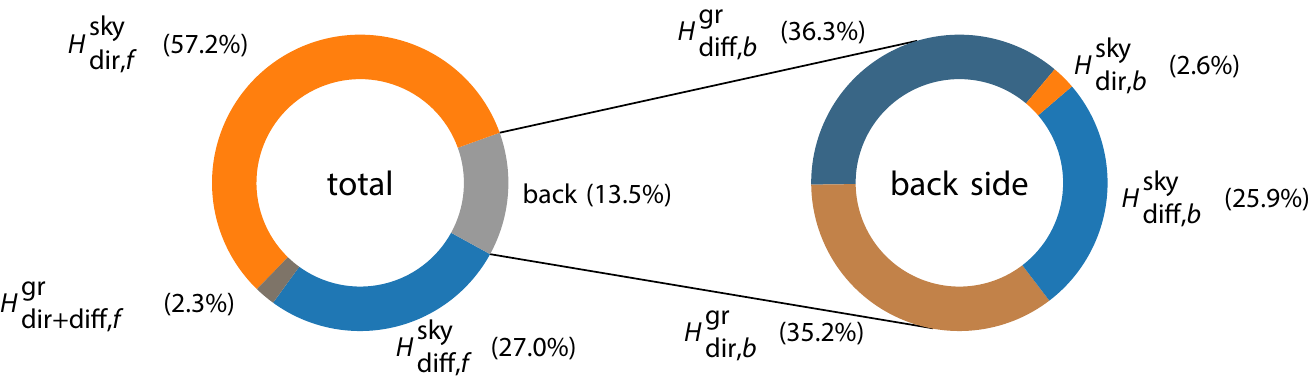}
 \caption{(left) Different annual radiant exposure components for a bifacial solar cell in Seattle.  (right) Detailed picture for the back side. Simulated with module spacing $d=10$~m, module tilt $\theta_m=42$\textdegree\ m, module height $h=0.5$~m and albedo $A=30\%$.}
 \label{fig:seattle-yield-contrib}
\end{figure*}

Figure \ref{fig:seattle-yield-contrib} shows how much the different irradiation components contribute to the total energy yield for a bifacial module with $d=10$~m module spacing and $\theta_m=42$\textdegree\ tilt in Seattle: About 69.5\% of the total energy arises from direct sunlight impinging onto the module front, 27.0\% are due to diffuse skylight impinging onto the front but the fraction of light that reaches the front from the ground is almost negligible. However, of the 13.5\% illumination onto the back, around 71.5\% arise from the ground.

\end{document}